\documentclass[sigconf]{acmart}
\AtBeginDocument{%
  }

\copyrightyear{2026}
\acmYear{2026}
\setcopyright{cc}
\setcctype{by-nc-nd}
\acmConference[FSE Companion '26]{34th ACM Joint European Software Engineering Conference and Symposium on the Foundations of Software Engineering}{July 05--09, 2026}{Montreal, QC, Canada}
\acmBooktitle{34th ACM Joint European Software Engineering Conference and Symposium on the Foundations of Software Engineering (FSE Companion '26), July 05--09, 2026, Montreal, QC, Canada}
\acmDOI{10.1145/3803437.3805573}
\acmISBN{979-8-4007-2636-1/2026/07}

\usepackage{balance}
\usepackage{xspace} 
\usepackage{multirow}
\usepackage{indentfirst}
\usepackage{xcolor}
\usepackage[most]{tcolorbox}

\newtcolorbox{rqbox}{
  colback=gray!8,
  colframe=black,
  boxrule=0.6pt,
  arc=1.5pt,
  left=6pt,
  right=6pt,
  top=6pt,
  bottom=6pt
}

\begin{document}

\title{From GUI Tests to Conversational Interaction: A New Perspective on App-Specific Voice Assistants}

%
\author{Xue Qin}
\email{xue.qin@villanova.edu}
\orcid{0000-0003-2503-1527}
\affiliation{%
  \institution{Villanova University}
  \city{Villanova}
  \state{Pennsylvania}
  \country{USA}
}

\author{Sumesh Surendran Letha}
\email{ssuren01@villanova.edu}
\affiliation{%
  \institution{Villanova University}
  \city{Villanova}
  \state{Pennsylvania}
  \country{USA}
}

\renewcommand{\shortauthors}{Xue et al.}

\begin{abstract}

Voice assistants are widely deployed on mobile platforms, yet most are designed as system-level services that remain poorly aligned with application-specific behavior. As a result, enabling voice interaction at the app level requires developers to manually reimplement application logic, leading to high development and maintenance costs.

We propose an LLM-driven approach to automating the development of app-specific voice assistants by repurposing GUI test code, which encodes behavior-preserving, executable specifications of application functionality. In this paper, we present a perspective in which large language models reinterpret GUI tests as bridges between application behavior and conversational interaction. By transforming test methods into app-specific VA artifacts, such as voice intents, capability descriptions, and executable action plans, our approach grounds voice assistants directly in existing application logic rather than external specifications.

We illustrate this vision through AppVA, a research prototype on Android. Our preliminary results across five open-source applications suggest that GUI test code can be reused beyond testing, enabling the synthesis of app-specific voice assistants and highlighting a broader research direction at the intersection of software testing, interaction design, and LLM-enabled automation.

\end{abstract}

\ccsdesc[500]{Software and its engineering~Software notations and tools}

\keywords{Voice Assistant, Test Code Reusing, Code Generation, Accessibility Development, LLM}

\maketitle

\section{Introduction}

Voice assistants (VA) are becoming a dominant interaction modality on mobile devices, with over 150 million users~\cite{emarketer2025voiceassistants} in the U.S. alone. As user expectations shift toward conversational interaction~\cite {destinationcrm2024voicebots}, developers face increasing pressure to integrate voice capabilities into millions of existing mobile applications.

A common industrial approach to adding voice-assistant (VA) capabilities to mobile applications relies on embedded, general-purpose VA services. Platforms such as Amazon Alexa require developers to reconstruct application functionality as external ``skills''~\cite{alexaSkillsKit} triggered by predefined voice commands. This approach forces developers to maintain a parallel interaction model that mirrors application logic, leading to growing divergence between voice workflows and application behavior as apps evolve. Prior studies report declining adoption and persistent challenges~\cite{arsAlexa2024, alexaInvocation2025, amazonAlexaErrors2019}, reflecting a deeper limitation: system-level VAs are not grounded in application-specific workflows.

In contrast, research in the Human–Computer Interaction (HCI) community has explored Voice User Interfaces (VUIs) for mobile apps. A VUI enables interaction through natural language commands, e.g., saying “start a workout” to trigger an application function. Recent Large Language Model (LLM)-based approaches improve user experience by supporting vocal task execution~\cite{song2024visiontasker}, but they treat voice interaction as a user-facing overlay. As a result, an app's capability logic is inferred from surface-level GUI context rather than its design and execution logic, making intent interpretation and action selection fragile and prone to mismatches~\cite{huang2025prompt2task}.

\textit{A key observation motivating our work is that modern mobile applications already contain executable, behavior-grounded specifications of user interaction in the form of GUI test code.} GUI tests encode valid sequences of user actions that reflect correct application behavior. They are authored by developers or testers with knowledge of the application’s logic and are continuously updated as the application evolves.
\textit{Rather than treating GUI tests solely as verification artifacts, we envision them as a reusable foundation for interaction design.}
By repurposing this existing test logic to support voice interaction, developers can avoid duplicating application behavior when adding voice assistant capabilities, thereby substantially reducing development and maintenance effort.
Recent advances in large language models make this vision attainable. 
By reinterpreting procedural test code into higher-level conversational artifacts, such as voice intents, capability descriptions, and executable action plans, LLMs enable a direct bridge between execution-level capabilities and voice-based interaction.

Building on this vision, we explore a new perspective on automating app-specific voice assistant development through GUI test code reusing. Specifically, we investigate how large language models can reinterpret executable test logic into voice assistant artifacts and enable app-specific voice interaction. This work supports a broader call to rethink how testing artifacts can be repurposed beyond the validation scope.
In particular, we make the following contributions:

\textbf{Approach}. 
We developed AppVA, a research prototype that explores how GUI test code can be transformed into app-specific voice assistant artifacts on Android. 
Through AppVA, we explore the range of voice assistant artifacts that can be derived through LLM-based interpretation and investigate the feasibility of this approach by demonstrating an end-to-end pipeline in which an LLM analyzes GUI test code and synthesizes conversational artifacts from test logic to final voice-ready specifications.
This work builds on prior efforts~\cite{weaver2024test2vareusingguitest} on reusing GUI test artifacts for voice assistant development, and introduces an LLM-based interpretation pipeline and richer conversational artifacts. 

\textbf{Evaluation}. 
Using AppVA as an exploratory research prototype, we examined how existing GUI test code can be repurposed to support app-specific voice interaction in practice. Across five open-source Android applications, we observed that AppVA transformed 19 out of 20 distinct GUI test methods into corresponding conversational artifacts, each representing an app-specific capability.
To further validate the feasibility of LLM-based interpretation, we evaluated how well test-derived intents aligned with user requests. When presented with 100 simulated voice queries, AppVA correctly mapped 88 queries to the intended capabilities across five apps. 

\textbf{Discussion and Insights}. 
Beyond quantitative outcomes, we reflect on the practical adoption of the AppVA for the developers and also discuss the possible future work for the researchers based on observed limitations.
These reflections highlight challenges and research opportunities at the intersection of software testing, interaction design, and LLM-driven automation.

\textbf{Open Science and Demonstration}. 
We make our research prototype publicly available and provide a live demonstration that illustrates how GUI test code can be transformed into executable, app-specific voice assistant artifacts in practice.~\footnote{
\url{https://nontestname.github.io/AppVA/}\\
\url{https://youtu.be/z4p19QL6ejw}
}

\vspace{-0.2cm}
\section{Related Work} \label{sec:related}

Recent work on voice assistants utilizes LLM agents to interpret GUI elements for task execution. This relies on LLMs interleaving reasoning with actions~\cite{wei2022chain, yao2022react}, learning autonomously~\cite{zhao2023expel}, and utilizing tools~\cite{schick2023toolformer}. Grounding these models requires translating semantics into feasible physical or digital plans~\cite{ahn2022do} via programmatic structures~\cite{singh2023progprompt}. In digital contexts, architectures bridging visual inputs with language priors~\cite{chen2022visualgpt} and GUI-specific visual models~\cite{hong2023cogagent} enable accurate interface perception. To manage interface complexity, systems leverage multi-agent debate to improve factuality~\cite{du2023improvingfactualityreasoninglanguage}, foster divergent thinking~\cite{liang2023encouraging}, and explore varied collaborative structures~\cite{zhang2023exploring}.
Despite these advances, challenges persist with unusual UI elements~\cite{song2024visiontasker}, complex parameterized tasks~\cite{vu2024gptvoicetasker}, or underspecified user instructions. Existing solutions typically expand training data or refine model architectures~\cite{huang2025prompt2task}. Moreover, inferring app behavior from surface-level GUI cues rather than code artifacts leads to implicit reasoning over incomplete interface cues, which may cause mismatches with intended semantics. Complementary to these approaches, our prior work~\cite{weaver2024test2vareusingguitest} explores the feasibility of transforming code from GUI test cases into app-specific voice-assistant capabilities without establishing the entire process. In this paper, we propose an LLM-based pipeline that defines and generates richer VA-ready artifacts, including capability descriptions, intent mappings, and action plans.

GUI test code reuse has been widely studied in software engineering, particularly in the context of improving testing efficiency and portability. Prior work has explored migrating test cases across mobile platforms, such as between iOS and Android~\cite{qin2019testmig}, as well as reusing tests across different applications~\cite{talebipour2021uitestmigration}. Other research has investigated repurposing existing test suites for related verification tasks, including configuration testing~\cite{fischer2019testreuse} and database-oriented testing~\cite{zhong2024testsuiteReuse}.
While these efforts demonstrate the versatility of test reuse within testing workflows, they remain focused on verification. The potential of GUI test code as a behavior-grounded specification for new interaction modalities, such as app-specific voice assistants, has received little attention. As a result, test code has not been explored as a semantic bridge between application logic and conversational interaction.

\vspace{-0.5cm}
\section{Approach} \label{sec:approach}

\subsection{Voice Assistant Artifacts}

To explore how to enable the app-specific voice assistants from existing test code, we first identify the core VA artifacts required to support conversational interaction and how they can be derived from GUI test logic using large language models.

\textbf{Grounded Interaction Units.}
At the foundation of our approach are \emph{grounded interaction units}, each corresponding to a single GUI test method written by developers or testers. 
These units encode validated sequences of user actions that accomplish a specific application task, including implicit preconditions and success criteria. As developer-authored and continuously maintained artifacts, grounded interaction units serve as a reliable source of truth for application behavior.

\textbf{VA Capability Definitions}
Building on grounded interaction units, we identify \emph{VA capability definitions} as the core behavioral abstractions exposed by an application’s voice assistant.
Each VA capability corresponds to a single GUI test method and represents an app-specific operation that can be invoked through voice interaction. 
VA capability definitions capture what the application can do in executable terms, providing a stable capability pool that evolves alongside the application.

\textbf{Capability Descriptions and Intents.}
To support voice interaction, VA capabilities must be accessible through natural language. We therefore identify \emph{capability descriptions and intents} as artifacts that connect VA capabilities with user utterances. Using LLM-based code interpretation and summarization, each VA capability is reinterpreted into a concise natural-language description that captures its purpose and effect, providing the semantic basis for intent recognition and mapping.

\textbf{Intent Summarization and Mappings.}
A voice assistant must not only recognize what users say, but also reason about how those requests relate to the application’s supported capabilities. We therefore identify \emph{intent summarization and mappings} as artifacts that mediate between user-level intents and capability-level intents. 
When a user request does not align with any supported capability, intent summarization enables the assistant to provide meaningful guidance by explaining what the application can do, rather than relying on GUI-based inference. 

\textbf{VA Action Plans.}
Finally, voice requests must be carried out through concrete application interactions. We therefore identify \emph{VA action} plans as the artifact that closes the loop from voice request to task execution. 
This plan represents an ordered sequence of GUI-level operations that execute a matched existing VA capability. By preserving the capability’s control flow and execution constraints, VA action plans ensure reliable and predictable task execution.

\begin{figure}[H]
\centering
\includegraphics[width=0.47\textwidth]{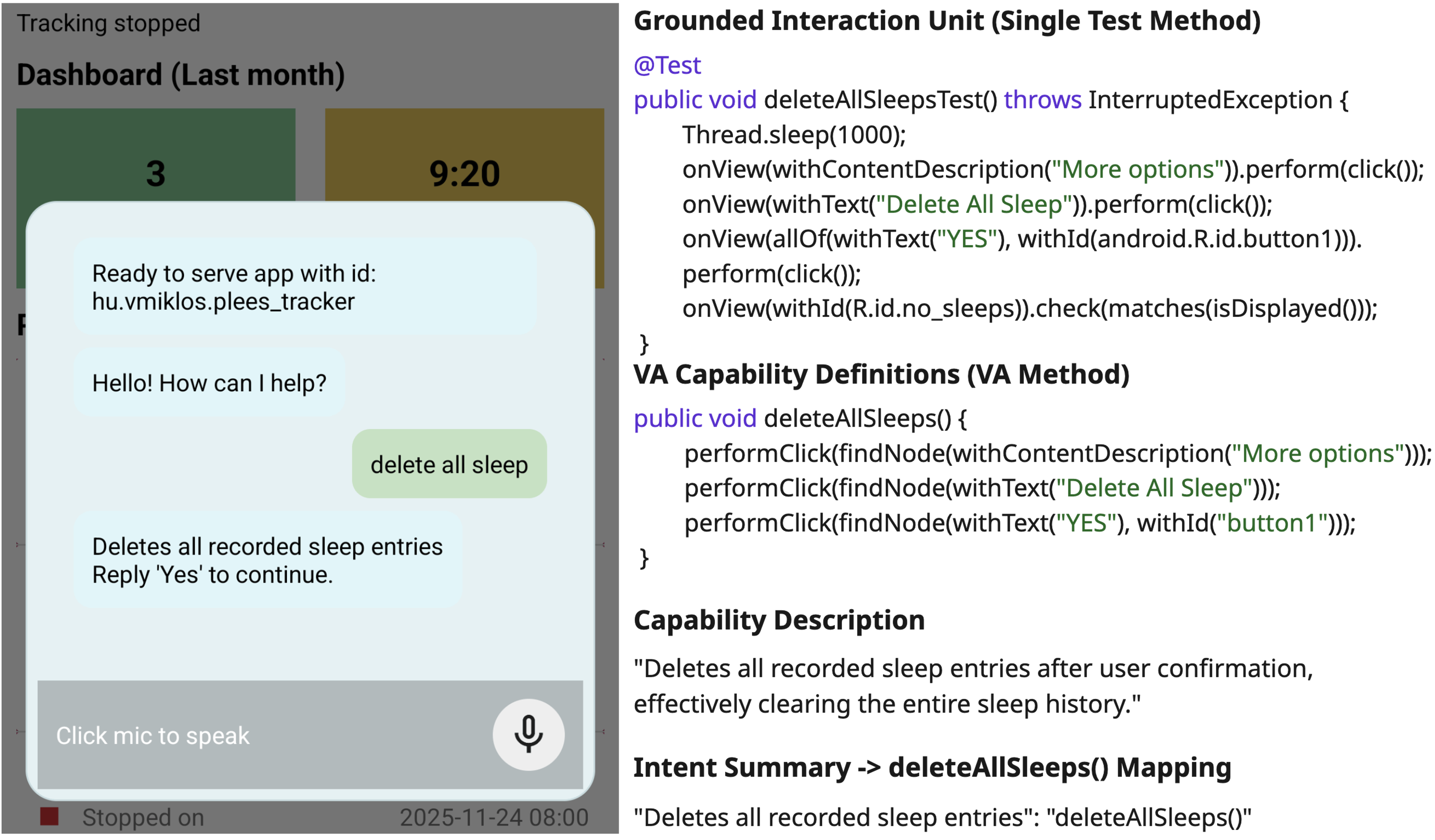}
\caption{Example of generated VA artifacts, including capability definition, description, and intent}
\vspace{-0.5cm}
\label{fig:example}
\end{figure}

Together, these artifacts enable a conceptual foundation for app-specific voice assistants whose conversational reasoning and execution remain aligned within the application's logic and behavior.
Figure~\ref{fig:example} illustrates how a GUI test method is transformed into corresponding VA artifacts in the proposed pipeline.

\begin{figure}[H]
\centering
\includegraphics[width=0.47\textwidth]{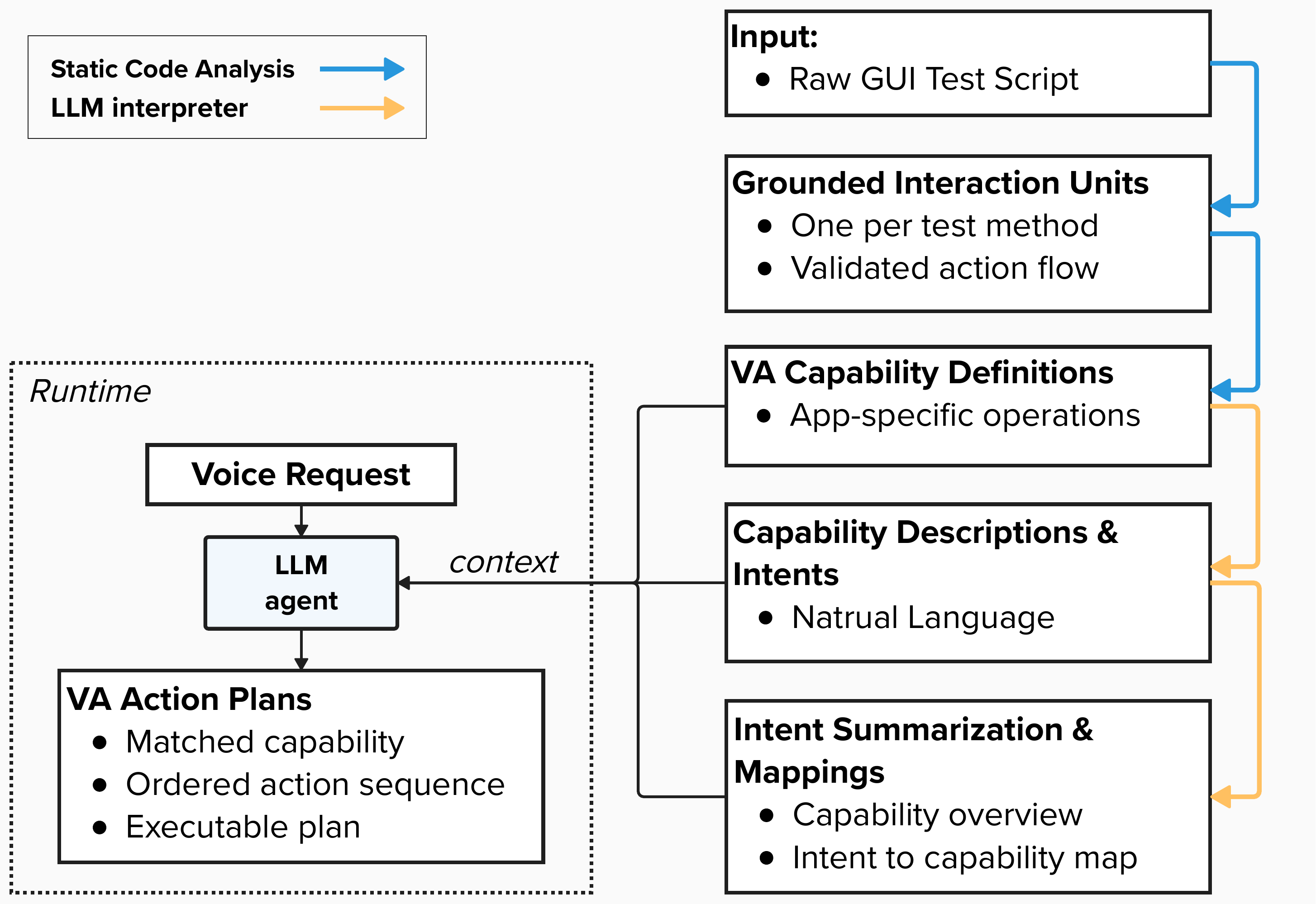}
\caption{Voice Assistant Artifacts Pipeline}
\label{fig:piepline}
\vspace{-0.5cm}
\end{figure}

\subsection{Conceptual VA Synthesis Pipeline}

\textbf{Static Pipeline.}
As illustrated in Figure~\ref{fig:piepline}, the pipeline begins with a static, pre-build phase that reinterprets developer-authored GUI test scripts into a structured set of voice assistant artifacts. GUI tests are treated as primary behavioral evidence because they encode how application functionality is exercised through validated interaction sequences.
Raw test scripts often contain multiple test methods as well as code that is unrelated to user interaction. 
The pipeline first applies static code analysis to isolate individual test methods. Each extracted method is treated as a grounded interaction unit. These units represent validated action flows for specific application tasks and preserve key execution properties such as the order of interactions.

Building on these units, the pipeline further abstracts framework-specific test logic into VA capability definitions. Each capability represents an app-specific operation that can be invoked through voice interaction, forming a per-application capability pool that reflects what the application can reliably do.
To make these capabilities accessible through natural language, the pipeline employs a GPT-4o-based interpreter that transforms extracted GUI test methods into structured artifacts and produces concise descriptions with corresponding intent candidates.
Beyond individual capabilities, the LLM constructs intent summarization and mappings, providing a structured overview of supported capabilities and connecting user-level requests to underlying operations. This also enables intent disambiguation, allowing the assistant to offer guidance when requests fall outside supported functionality. Importantly, this reasoning remains grounded in validated application behavior rather than inferred solely from GUI context.

\textbf{Runtime Pipeline.}
At interaction time, when a user issues a voice request, an LLM agent will interpret the request using pre-generated capability descriptions, intent summaries, and mappings as context. 
If a capability matches the request, the agent retrieves or synthesizes a corresponding VA action plan, which specifies an ordered sequence of GUI-level operations to perform the requested task on the application. 
In this way, the assistant closes the loop from conversational intent to task execution. 
If the request is ambiguous or unsupported, the agent provides clarification or guidance to the user based on the intent summarization.

In this conceptual pipeline, static code analysis extracts execution structure from test code, while LLM-based interpretation performs semantic lifting by translating execution logic into natural language and enabling conversational interaction.

\textbf{Example.}
As shown in Figure~\ref{fig:example}, the pipeline first extracts a grounded interaction unit from the raw GUI test script and reinterprets it as a VA capability definition, preserving only interaction-relevant logic. The LLM then summarizes this capability into a natural-language description and intent.

\section{Proof of Concept}

\subsection{Research Questions}

We focus on the following research questions:

\textbf{RQ1: }\textit{How effectively can the static pipeline repurpose GUI test code to synthesize executable, app-specific voice assistant capabilities?}
This research question examines whether GUI test methods can be transformed into voice assistant capabilities that correctly represent application behavior and support execution.

\textbf{RQ2: }\textit{How effectively can the runtime pipeline align simulated real-world voice requests with app-specific capabilities under varying levels of ambiguity?}
This research question evaluates whether LLM-derived capability descriptions, intent summaries, and mappings enable accurate and robust intent–capability alignment when user requests are underspecified or ambiguous.

\begin{table}[H]
\centering
\caption{Number of successful requests and accuracy level}
\label{tab:eval-intent}
\begin{tabular}{l|ccccc}
\hline
\multicolumn{1}{c|}{\multirow{2}{*}{\textbf{App Id}}} & \multicolumn{5}{c}{\textbf{Request Accuracy Level}}                                                                       \\ \cline{2-6} 
\multicolumn{1}{c|}{}                        & \multicolumn{1}{c|}{\textbf{0.0}} & \multicolumn{1}{c|}{\textbf{0.25}} & \multicolumn{1}{c|}{\textbf{0.5}} & \multicolumn{1}{c|}{\textbf{0.75}} & \textbf{1.0} \\ \hline
com.faltenreich.diaguard                      & \multicolumn{1}{c|}{4}   & \multicolumn{1}{c|}{4}    & \multicolumn{1}{c|}{4}   & \multicolumn{1}{c|}{4}    & 4   \\ \hline
com.flauschcode.broccoli                      & \multicolumn{1}{c|}{3}   & \multicolumn{1}{c|}{2}    & \multicolumn{1}{c|}{2}   & \multicolumn{1}{c|}{1}    & 1   \\ \hline
com.futsch1.medtimer                          & \multicolumn{1}{c|}{4}   & \multicolumn{1}{c|}{4}    & \multicolumn{1}{c|}{4}   & \multicolumn{1}{c|}{4}    & 3   \\ \hline
hu.vmiklos.plees\_tracker                     & \multicolumn{1}{c|}{4}   & \multicolumn{1}{c|}{4}    & \multicolumn{1}{c|}{4}   & \multicolumn{1}{c|}{4}    & 4   \\ \hline
org.totschnig.myexpenses                      & \multicolumn{1}{c|}{4}   & \multicolumn{1}{c|}{4}    & \multicolumn{1}{c|}{4}   & \multicolumn{1}{c|}{4}    & 4   \\ \hline
Total                                         & \multicolumn{1}{c|}{19}  & \multicolumn{1}{c|}{18}   & \multicolumn{1}{c|}{18}  & \multicolumn{1}{c|}{17}   & 16  \\ \hline
\end{tabular}

\end{table}

\vspace{-0.5cm}
\subsection{Applications and Evaluation Setup}

We evaluated our approach using five open-source Android applications from diverse domains (e.g., health tracking and personal finance) to cover common interaction patterns such as navigation, data entry, and record management. For each application, we selected four representative GUI test methods written in Espresso, covering typical user tasks such as navigation, data entry, and configuration, yielding 20 test methods in total. While limited in scale, this selection is appropriate for demonstrating feasibility across varied app behaviors.
To assess intent–capability alignment, we generated synthetic natural-language voice requests using GPT-4o; real user evaluation is left for future work. For each test method, five requests with varying specificity (from detailed to vague) were synthesized, from detailed (1.0) to vague (0.0), yielding 100 requests in total. Each request was labeled with its ground-truth capability.

\subsection{Findings}

\subsubsection{Static Pipeline Feasibility}
First, we examine whether the static pipeline can repurpose GUI test code to synthesize executable, app-specific voice assistant capabilities. Across the five evaluated applications, the pipeline generated voice assistant artifacts for all 20 selected GUI test methods, of which 19 were correctly triggered and executed using appropriate voice requests.
The only failure occurred in the seasonal recipe application for the capability that configures the seasonal calendar. In this case, the LLM misinterpreted the test logic during capability abstraction, resulting in an incorrect capability description and a subsequent execution failure. This example illustrates that static capability synthesis is sensitive to semantic interpretation, especially for configuration-heavy behaviors.
\textbf{Answer to RQ1.}
\textit{GUI test code can be repurposed into executable, app-specific voice assistant
capabilities, with 19 of 20 capabilities executing successfully.}

\subsubsection{Runtime Pipeline Feasibility}
Next, we examine whether the runtime pipeline can align simulated voice requests with varying levels of ambiguity. The left screen in Figure~\ref{fig:example} shows a runtime example of AppVA for app \texttt{plees\_tracker}, where the generated VA artifacts match a user request of ``delete all sleep'' to \texttt{deleteAllSleep()} capability. 
A match is deemed correct when the pipeline selects the ground-truth capability associated with the test method used to generate the request, without human evaluation.
Across the five applications, 88 out of 100 synthesized voice requests were correctly matched to their intended capabilities. Three out of five applications achieved perfect matching across all request specificity levels.
As shown in Table~\ref{tab:eval-intent}, most mismatches occurred in the recipe application (\texttt{broccoli}), whose capabilities require multiple detailed parameters. In these cases, vague or underspecified requests increased ambiguity and made precise intent–capability alignment more challenging. These observations suggest that the runtime pipeline is generally effective for intent resolution, while highlighting limitations when capability semantics depend on fine-grained argument details. The detailed evaluation report can be found in the paper's website.
\textbf{Answer to RQ2.}
\textit{The runtime pipeline correctly matched 88\% of the simulated voice requests.}

\vspace{-0.3cm}
\section{Discussion and Future Work}\label{sec:future_discussion}

\textbf{Implications for Developers.}
Our findings suggest that GUI test code can serve as a reusable, behavior-grounded specification for app-specific voice assistants. By repurposing validated test logic, developers can expose application functionality through voice interaction without reimplementing behavior or designing workflows from scratch. Because test suites evolve with applications, voice capabilities derived from tests can remain aligned with application behavior, reducing development and maintenance effort. More broadly, this approach shifts voice assistant design from GUI-centric inference to behavior-centric grounding, mitigating semantic drift in complex applications.

\textbf{Design Insights and Limitations.}
Our exploration reveals several design insights and limitations. First, semantic misinterpretation during the static pipeline remains a risk; in one case, the LLM mischaracterized a configuration-oriented test, leading to an incorrect capability and execution failure. Second, the runtime pipeline is sensitive to ambiguity in user requests. While test-derived descriptions support intent matching, vague or underspecified requests remain challenging when capabilities share overlapping semantics. Finally, capabilities are tied to tests that validate limited behaviors. As our goal is to demonstrate feasibility rather than assess test coverage, the approach depends on available test suites and may be limited when application behavior is only partially covered. We do not evaluate scenarios where tests evolve or become misaligned with application changes, nor do we analyze LLM-related trade-offs such as cost, latency, and output variability. Overall, this rigidity reduces flexibility for novel or partially specified requests.

\textbf{Future Research Directions}
These limitations suggest several directions for future research. Improving semantic accuracy during capability abstraction remains critical, especially for configuration-heavy or implicit logic; hybrid approaches combining LLMs with lightweight static analysis or developer guidance may help. At runtime, more effective clarification strategies are needed to resolve ambiguous requests. 
In addition, parameter generalization (e.g., deriving inputs from test values or UI fields) could enable capabilities to handle a wider range of inputs while preserving correctness.

\vspace{-0.2cm}
\section{Conclusion}

In summary, this work explores how GUI test code can be repurposed as a behavior-grounded foundation for app-specific voice assistants using LLM-based interpretation. Our results show that many existing tests can be transformed into executable voice capabilities and support simulated voice requests with promising accuracy. While challenges remain,
this study highlights the potential of reusing test artifacts to reduce the cost of building app-specific voice assistants.

\balance
\bibliographystyle{ACM-Reference-Format}
\bibliography{sample-base}

\end{document}